\documentclass[aps,prd,twocolumn,superscriptaddress,showkeys,nofootinbib]{revtex4-2}

\usepackage[utf8]{inputenc}
\usepackage{amsmath}
\usepackage{graphicx}
\usepackage{xcolor}

\def\fs#1{\setbox0=\hbox{$#1$}#1\hskip-\wd0\dimen0=5pt\advance\dimen0
by-\ht0\advance\dimen0 by\dp0\lower0.5\dimen0\hbox to\wd0{\hss\sl/\/\hss}}

\begin{document}

\title{Phenomenological mass model for exotic hadrons and predictions for masses of non-strange dibaryons as hexaquarks}

\author{Christoffer Beiming}
\email[]{cbeiming@kth.se}
\affiliation{Department of Physics, School of Engineering Sciences, KTH Royal Institute of Technology, AlbaNova University Center, Roslagstullsbacken 21, SE--106~91 Stockholm, Sweden}

\author{Jesper Gr{\"o}nroos}
\email[]{jesgro@kth.se}
\affiliation{Department of Physics, School of Engineering Sciences, KTH Royal Institute of Technology, AlbaNova University Center, Roslagstullsbacken 21, SE--106~91 Stockholm, Sweden}

\author{Tommy Ohlsson}
\email[]{tohlsson@kth.se}
\affiliation{Department of Physics, School of Engineering Sciences, KTH Royal Institute of Technology, AlbaNova University Center, Roslagstullsbacken 21, SE--106~91 Stockholm, Sweden}
\affiliation{The Oskar Klein Centre for Cosmoparticle Physics, AlbaNova University Center, Roslagstullsbacken 21, SE--106 91 Stockholm, Sweden}

\begin{abstract}
We investigate the mass spectra of exotic hadrons known as hexaquarks in the form of dibaryons. We use a phenomenological model based on an extended version of the G{\"u}rsey--Radicati mass formula for hadrons to include non-charmed baryons, charmed baryons, and non-strange dibaryons to be able to predict masses of potential dibaryon states. We perform six numerical fits of this model to input data for three different sets of masses of baryons and dibaryons. We find that the model can fit some of the data sets well, especially the sets including charmed baryons and non-strange dibaryons, and observe that the predicted mass of one of the dibaryons is close to the measured mass of the observed hexaquark candidate $d^*(2380)$ reported by the WASA-at-COSY experiment. The predicted mass of the deuteron is slightly larger than its measured mass. Finally, for the data sets including charmed baryon and non-strange dibaryon masses, we find that the predicted masses of potential dibaryon states are all in the range from 1900~MeV to 3700~MeV.
\end{abstract}

\maketitle

\section{Introduction\label{sec:intro}}

Quarks are subatomic particles that were proposed independently by Gell-Mann \cite{GellMann:1964nj} and Zweig \cite{Zweig:1964jf} in 1964, and verified by experiments a few years later. They are never found isolated in Nature, but rather exist in composite structures called hadrons. Observed hadrons normally consist of quark-antiquark pairs or collections of three quarks (or three antiquarks). However, apart from the constraint of hadrons having zero net color, the quark model does not impose any limitation on the number of quarks that can constitute a hadron. The existence of so-called exotic hadrons consisting of more than three quarks was also proposed by Gell-Mann \cite{GellMann:1964nj} in 1964, and such hadrons have been observed by various particle accelerator experiments in recent years \cite{Jaffe:2004ph,Cowan:2018fki}. The most commonly discussed exotic hadrons are those composed of four, five or six quarks, known as tetraquarks, pentaquarks and hexaquarks, respectively (see e.g.~Refs.~\cite{Esposito:2016noz,Ali:2019roi}).

The proposition of six quarks combined into a single structure, although not necessarily called a hexaquark, was initially proposed by Dyson and Xuong \cite{Dyson:1964xwa} in 1964, as a so-called dibaryon state. They predicted non-strange S-wave dibaryon states $D_{IJ}$, where $I$ and $J$ denote isospin and spin, respectively. The proposal was that the deuteron, $D_{01}(NN)$, and a virtual state, $D_{10}(NN)$, were contained in the $\overline{{\bf 10}}$ and ${\bf 27}$ representations of the group SU(3) that is part of the decomposition ${\rm SU(3)} \times {\rm SU(2)} \subset {\rm SU(6)}$. They also predicted four additional states, $D_{03}(\Delta\Delta)$, $D_{30}(\Delta\Delta)$, $D_{12}(N\Delta)$, and $D_{21}(N\Delta)$, based on group theoretical symmetry arguments. Yet another hexaquark was proposed by Jaffe in 1976 \cite{Jaffe:1976yi}. It was assumed to have the quark content $uuddss$ and given the name dihyperon or the H-particle. Note that the name hyperon is nowadays normally used for baryons containing one or more strange quarks. The stability of the H-particle and many other dibaryon candidates was later studied further in several works by Leandri and Silvestre-Brac, see Ref.~\cite{Leandri:1997ge} and references therein. Also, in Refs.~\cite{Bashkanov:2013cla,Brodsky:2015wza,Maiani:2015iaa}, the ideas of hidden-color dibaryons and hidden-charm hexaquarks were discussed. For the pure hexaquark sector, simply meaning six quarks and no antiquarks, a large systematic study of dibaryon candidates was carried out \cite{Leandri:1997ge} by means of a schematic chromomagnetic model (see e.g.~Refs.~\cite{Hogaasen:2005jv,Abud:2009rk,An:2020vku}). Results pointed to dibaryon candidates most favorable for stability being composed of three different pairs of identical quarks and  these states were also studied in Ref.~\cite{Leandri:1997ge}. Despite the possibility of such resonances not being excluded, the investigations entailed that stable dibaryon states are not favored -- there always exists a two-baryon channel with lower energy. However, the search for hexaquarks revived in 2014, when the WASA-at-COSY collaboration revealed observations of a narrow resonance-like structure around 2380~MeV by studying the double-pionic fusion channels $pn \to d \pi^0 \pi^0$ and $pn \to d \pi^+ \pi^-$. This structure was named $d^*(2380)$ and data suggested it having quantum numbers $I(J^P) = 0(3^+)$ \cite{Adlarson:2014pxj}. A serious candidate for the $d^*(2380)$ has later been thought to be the $D_{03}$ state, proposed fifty years earlier, having the proper quantum numbers and a predicted mass of about 2350~MeV. More recently, Huang {\it et al.}~\cite{Huang:2018izl} investigated the six dibaryon candidates proposed by Dyson and Xuong, using dynamical calculations and a simple mass formula previously applied to baryons and pentaquarks. In particular, they investigated the possibility of the $D_{21}$ as a bound-state dibaryon candidate and concluded that models that obtain the experimental $d^*(2380)$ as the $D_{03}$ are not compatible with also obtaining the $D_{21}$ as a bound state. Hexaquarks have also been studied in chiral quark models \cite{Huang:2014kja,Lu:2017uey}, using QCD sum rules~\cite{Wang:2017sto,Wang:2020fuh,Wang:2021isu}, and by constructing wave functions with effective potentials \cite{Kim:2020rwn}. Finally, note that alternative interpretations of the observed $d^*(2380)$ structure have been proposed, see e.g.~Refs.~\cite{Ikeno:2021frl,Molina:2021bwp}.

In this work, we will consider the simple mass formula for hadrons discussed in Ref.~\cite{Huang:2018izl}, which could predict the masses of hexaquarks using numerical fits to measured masses of various baryons. In general, we will use an extended version of this formula, fit it to larger data sets of baryons that have not been considered previously, and predict masses of the six non-strange S-wave baryon states $D_{IJ}$ that could constitute hexaquarks.

This work is organized as follows. In Sec.~\ref{sec:model}, we present the phenomenological mass model based on the so-called extended G{\"u}rsey--Radicati mass formula for hadrons that can predict masses of hexaquarks, and describe the numerical fitting procedure. Then, in Sec.~\ref{sec:results}, we introduce three different data sets for measured masses of baryons, perform six numerical fits of this model, and state and discuss the results of these fits in detail. The predicted masses of hexaquarks are also presented. Finally, in Sec.~\ref{sec:summary}, we summarize our main results and conclude based on our stated results.

\section{Model and fitting procedure}
\label{sec:model}

\subsection{Extended G{\"u}rsey--Radicati model}

Already in 1964, G{\"u}rsey and Radicati proposed a simple mass formula for baryons, based on the breaking of ${\rm SU(6)} \supset {\rm SU(2)}_s \times {\rm SU(3)}_f$ spin-flavor symmetry~\cite{Gursey:1992dc}. As pointed out in Ref.~\cite{Giannini:2005ks}, the G{\"u}rsey--Radicati (GR) mass formula describes quite well the way in which symmetry breaking affects the mass spectrum of baryons, despite its simplicity. In the original work of G{\"u}rsey and Radicati, the mass formula is given by (up to differences in notation)
\begin{equation}
M = M_0 + A J (J + 1) + B Y + C \left[ I (I + 1) - \frac{1}{4} Y^2 \right] \,,
\label{eq:GR}
\end{equation}
where $J$, $Y$, and $I$ denote spin, hypercharge, and isospin, respectively, and $M_0$, $A$, $B$, and $C$ are phenomenological model parameters. These parameters are obtained from fits to baryon data. An extension to Eq.~(\ref{eq:GR}) was introduced and studied in Ref.~\cite{Santopinto:2016pkp}, and further investigated in Ref.~\cite{Holma:2019lxe}, where applications to pentaquark systems are considered. The extended version of the GR mass formula reads \cite{Santopinto:2016pkp}
\begin{align}
M' &= M_0 + A J (J + 1) + B Y + C \left[ I (I + 1) - \frac{1}{4} Y^2 \right] \nonumber\\
&+ D {\cal C}_2 + E N_c \,,
\label{eq:GRext}
\end{align}
where ${\cal C}_2$ is the eigenvalue of the ${\rm SU(3)}_f$ Casimir operator, $N_c$ is the number of constituent charm quarks in the considered baryon, and $D$ and $E$ are two additional model parameters. In Refs.~\cite{Santopinto:2016pkp,Holma:2019lxe}, the parameters are fixed using the baryon spectrum and pentaquark masses are then calculated under the assumption that these parameter values are universally valid beyond baryons. Furthermore, the extended GR formula~(\ref{eq:GRext}) is used in the context of hexaquarks in Ref.~\cite{Huang:2018izl}, when studying the possible dibaryon candidates $D_{IJ}$ suggested by Dyson and Xuong in 1964. For these dibaryon states, a simple formula for the eigenvalue of the Casimir operator is presented as \cite{Dyson:1964xwa}
\begin{equation}
{\cal C}_2 = 12 + 2 I (I + 1) \,.
\label{eq:C2}
\end{equation}

\subsection{Fitting procedure}

Now, we continue to the fitting procedure of baryon data. Following the approach of Huang {\it et al.}~\cite{Huang:2018izl}, we use the extended GR mass formula~(\ref{eq:GRext}) to calculate the mass spectrum of dibaryons previously predicted by Dyson and Xuong~\cite{Dyson:1964xwa}, imposing universality of model parameters.

The six model parameters are fitted to experimental data on hadron masses, retrieved from the Particle Data Group (PDG)~\cite{Zyla:2020zbs} of 2020. In Ref.~\cite{Huang:2018izl}, the data set used consists of eight non-charmed baryons and six non-strange dibaryons, where of the latter only two have experimentally measured and verified masses. To distinguish our study from theirs, we additionally include eight charmed baryons, and furthermore, we only use the two verified dibaryon candidates $NN$ and $d^*(2380)$ for $D_{01}$ and $D_{03}$, respectively. In Tab.~\ref{tab:baryon}, the baryons along with their corresponding masses and quantum numbers are presented, and likewise in Tab.~\ref{tab:dibaryon}, the corresponding properties of the dibaryon candidates are found.

\begin{table}
\centering
\caption{Experimental values and errors for the masses of 16 selected baryons including corresponding quantum numbers. All values are taken from Ref.~\cite{Zyla:2020zbs}. The 16 selected baryons are all ground-state non-strange baryons, hyperons, or charmed baryons.}
\begin{tabular}{l c c c c c c c c}
\hline\hline
Baryon & Exp.~mass [MeV] & Exp.~error [MeV] & $J$ & $Y$ & $I$ & ${\cal C}_2$ & $N_c$ \\ [0.1ex] 
\hline \\[-2.3ex]
$N$ & 939.565413 & $\pm 6 \cdot 10^{-6}$ & $\frac{1}{2}$ & 1 & $\frac{1}{2}$ & 3 & 0 \\ [0.3ex] 
$\Lambda^0$ & 1115.683 & $\pm 0.006$ & $\frac{1}{2}$ & 0 & 0 & 3 & 0 \\ [0.3ex]
$\Sigma^0$ & 1192.642 & $\pm 0.024$ & $\frac{1}{2}$ & 0 & 1 & 3 & 0 \\ [0.3ex]
$\Xi^0$ & 1314.86 & $\pm 0.20$ & $\frac{1}{2}$ & $-1$ & $\frac{1}{2}$ & 3 & 0 \\ [0.3ex]
$\Delta^0$ & 1232 & $\pm 2$ & $\frac{3}{2}$ & 1 & $\frac{3}{2}$ & 6 & 0 \\ [0.3ex]
$\Sigma^{*0}$ & 1383.7 & $\pm 1.0$ & $\frac{3}{2}$ & 0 & 1 & 6 & 0 \\ [0.3ex]
$\Xi^{*0}$ & 1531.80 & $\pm 0.32$ & $\frac{3}{2}$ & $-1$ & $\frac{1}{2}$ & 6 & 0 \\ [0.3ex]
$\Omega^-$ & 1672.45 & $\pm 0.29$ & $\frac{3}{2}$ & $-2$ & 0 & 6 & 0 \\ [0.3ex]
\hline \\[-2.3ex]
$\Lambda_c^+$ & 2286.46 & $\pm 0.14$ & $\frac{1}{2}$ & $\frac{2}{3}$ & 0 & $\frac{4}{3}$ & 1 \\ [0.3ex]
$\Sigma_c^0$ & 2453.75 & $\pm 0.14$ & $\frac{1}{2}$ & $\frac{2}{3}$ & 1 & $\frac{10}{3}$ & 1 \\ [0.3ex]
$\Xi_c^0$ & 2470.90 & ${}_{-0.29}^{+0.22}$ & $\frac{1}{2}$ & $-\frac{1}{3}$ & $\frac{1}{2}$ & $\frac{4}{3}$ & 1 \\ [0.3ex]
$\Xi_c^{\prime 0}$ & 2579.2 & $\pm 0.5$ & $\frac{1}{2}$ & $-\frac{1}{3}$ & $\frac{1}{2}$ & $\frac{10}{3}$ & 1 \\ [0.3ex]
$\Omega_c^0$ & 2695.2 & $\pm 1.7$ & $\frac{1}{2}$ & $-\frac{4}{3}$ & 0 & $\frac{10}{3}$ & 1 \\ [0.3ex]
$\Omega_c^{*0}$ & 2765.9 & $\pm 2.0$ & $\frac{3}{2}$ & $-\frac{4}{3}$ & 0 & $\frac{10}{3}$ & 1 \\ [0.3ex]
$\Sigma_c^{*0}$ & 2518.48 & $\pm 0.20$ & $\frac{3}{2}$ & $\frac{2}{3}$ & 1 & $\frac{10}{3}$ & 1 \\ [0.3ex]
$\Xi_c^{*0}$ & 2646.38 & ${}_{-0.23}^{+0.20}$ & $\frac{3}{2}$ & $-\frac{1}{3}$ & $\frac{1}{2}$ & $\frac{10}{3}$ &1 \\ [0.3ex]
\hline\hline
\end{tabular}
\label{tab:baryon}
\end{table}

\begin{table}
\centering
\caption{Experimental values and errors for the masses of dibaryons. The values for $D_{01}(NN)$ [i.e.~$NN$] and $D_{03}(\Delta\Delta)$ [i.e.~$d^*(2380)$] are adopted from Refs.~\cite{Adlarson:2014pxj,Zyla:2020zbs}. Dibaryon quantum numbers are as used in Ref.~\cite{Huang:2018izl}.}
\begin{tabular}{l c c c c c c c c}
\hline\hline
Dibaryon & Exp.~mass [MeV] & Exp.~error [MeV] & $J$ & $Y$ & $I$ & ${\cal C}_2$ & $N_c$ \\ [0.1ex] 
\hline \\[-2.3ex]
$D_{01}(NN)$ & 1875.612943 & $\pm 10^{-6}$ & 1 & 2 & 0 & 12 & 0 \\ [0.3ex] 
$D_{10}(NN)$ & $-$ & $-$ & 0 & 2 & 1 & 16 & 0 \\ [0.3ex]
$D_{03}(\Delta\Delta)$ & 2380 & $\pm 10$ & 3 & 2 & 0 & 12 & 0 \\ [0.3ex]
$D_{30}(\Delta\Delta)$ & $-$ & $-$ & 0 & 2 & 3 & 36 & 0 \\ [0.3ex]
$D_{12}(N\Delta)$ & $-$ & $-$ & 2 & 2 & 1 & 16 & 0 \\ [0.3ex]
$D_{21}(N\Delta)$ & $-$ & $-$ & 1 & 2 & 2 & 24 & 0 \\ [0.3ex]
\hline\hline
\end{tabular}
\label{tab:dibaryon}
\end{table}

These data are divided into three sets, for each of which separate fits and calculations will be carried out. The division of the data is as follows:
\begin{itemize}
\item[Set ${\bf I}$:] Data for the eight non-charmed baryons in the upper half of Tab~\ref{tab:baryon}.
\item[Set ${\bf II}$:] Data for the eight non-charmed baryons and the eight charmed baryons in Tab~\ref{tab:baryon}.
\item[Set ${\bf III}$:] Data for the eight non-charmed baryons, the eight charmed baryons, and the two dibaryon candidates in Tabs.~\ref{tab:baryon} and \ref{tab:dibaryon}.
\end{itemize}
The parameters for each data set are then determined by minimizing the $\chi^2$-function
\begin{equation}
\chi^2 = \sum_i \left( \frac{M_i^{\rm exp.} - M'_i}{\sigma_i^{\rm exp.}} \right)^2 \,,
\label{eq:chi2}
\end{equation}
where $M_i^{\rm exp.}$ are the tabulated experimental masses, $\sigma_i^{\rm exp.}$ the corresponding experimental uncertainties, and $M'_i$ the input masses for data sets ${\bf I}$, ${\bf II}$, and ${\bf III}$, respectively.

The minimization of the $\chi^2$-function~(\ref{eq:chi2}) is performed using both the built-in function {\tt Minimize} in Mathematica \cite{reference.wolfram_2021_minimize} and a numerical method called basin-hopping, which is an iterative technique for globally optimizing a scalar function of one or several variables. Basin-hopping is a two-phase method consisting of a stochastic global stepping algorithm and local minimization at each step \cite{reference.scipy.optimize.basinhopping}, which exist as built-in functions in the SciPy library. For the minimization, several numerical methods can be used, but in this work the above-mentioned function in Mathematica and the so-called BFGS-algorithm are chosen.

\section{Numerical fits and results}
\label{sec:results}

\subsection{Data sets}

After having divided the data into three sets, we fit the parameters of the extended GR mass formula~(\ref{eq:GRext}) with Eq.~(\ref{eq:C2}) in two ways for each set. Firstly, we use the experimental uncertainties of the hadron masses -- these fits are called unprimed and denoted ${\bf I}$, ${\bf II}$, and ${\bf III}$. Secondly, we do what is called a ``1~\% error fit'', i.e.~setting each uncertainty $\sigma_i$ to one percent of the corresponding experimentally determined mass. This can sometimes yield a better fit, although the precision of the results cannot be better than just 1~\%. These fits are called primed and denoted ${\bf I}'$, ${\bf II}'$, and ${\bf III}'$.

\subsection{Numerical fits and results}

In Tab.~\ref{tab:fitparam}, resulting parameter values for each case are presented along with the value of the minimized $\chi^2$-function~(\ref{eq:chi2}). We then use the GR formula~(\ref{eq:GRext}) with these parameters together with quantum numbers from Tab.~\ref{tab:dibaryon}, to predict the mass spectrum of the dibaryon candidates. The predictions for the dibaryon mass spectrum are presented in Tab.~\ref{tab:dibaryonres}. For fits ${\bf III}$ and ${\bf III'}$, the experimental masses of $D_{01}$ and $D_{03}$ are used as inputs, hence they are not included as predictions, whereas the resulting predictions for the masses of the other four dibaryons, i.e.~$D_{10}$, $D_{30}$, $D_{12}$, and $D_{21}$, are all found to be in the rather narrow spectrum of $(1900,2700)$~MeV. In fact, for fits~${\bf II}$--${\bf III}'$, it turns out that the predictions for the masses of all six dibaryons lay in the interval $(1900,2900)$~MeV, except for the predicted mass of $D_{30}$ for fits~${\bf II}$ and ${\bf II}'$ that is slightly larger.

\begin{table}
\centering
\caption{Fitted free parameter values for the six different fits and the corresponding values of the $\chi^2$-function. All values are presented with four significant figures. It should be noted that the results of fits~${\bf II}$ and ${\bf II}'$ basically correspond to updated results of fits~I and III in Ref.~\cite{Holma:2019lxe}.}
\begin{tabular}{l r r r r r r}
\hline\hline
Parameter & Fit ${\bf I}$ & Fit ${\bf I}'$ & Fit ${\bf II}$ & Fit ${\bf II}'$ & Fit ${\bf III}$ & Fit ${\bf III}'$ \\ [0.1ex] 
\hline \\[-2.3ex]
$M_0$ [MeV] & 280.1 & 280.9 & 954.7 & 964.5 & 1057 & 1029 \\ [0.3ex]
$A$ [MeV] & $-281.4$ & $-282.7$ & 18.69 & 22.43 & 42.18 & 46.63 \\ [0.3ex]
$B$ [MeV] & $-195.3$ & $-193.8$ & $-195.3$ & $-189.0$ & $-195.7$ & $-198.7$ \\ [0.3ex]
$C$ [MeV] & 38.43 & 33.00 & 38.38 & 33.03 & 39.30 & 40.81 \\ [0.3ex]
$D$ [MeV] & 348.9 & 349.2 & 48.96 & 44.76 & 8.998 & 15.69 \\ [0.3ex]
$E$ [MeV] & $\times$ & $\times$ & 1377 & 1360 & 1339 & 1326 \\ [0.3ex]
\hline \\[-2.3ex]
$\chi^2$ & 6248 & 1.925 & 37820 & 8.490 & 312300 & 26.16 \\ [0.3ex]
$\chi^2/{\rm d.o.f.}$ & 2083 & 0.6417 & 3782 & 0.8490 & 26025 & 2.180 \\ [0.3ex]
\hline\hline
\end{tabular}
\label{tab:fitparam}
\end{table}

\begin{table}
\centering
\caption{Predicted mass values of dibaryons for the six different fits. All values in this table are presented in units of MeV and with four significant digits.}
\begin{tabular}{l r r r r r r}
\hline\hline
Dibaryon & Fit ${\bf I}$ & Fit ${\bf I}'$ & Fit ${\bf II}$ & Fit ${\bf II}'$ & Fit ${\bf III}$ & Fit ${\bf III}'$ \\ [0.1ex] 
\hline \\[-2.3ex]
$D_{01}(NN)$ & 3755 & 3766 & 2106 & 2100 & $\times$ & $\times$ \\ [0.3ex]
$D_{10}(NN)$ & 5790 & 5795 & 2341 & 2300 & 1906 & 1953 \\ [0.3ex]
$D_{03}(\Delta\Delta)$ & 940.9 & 939.4 & 2292 & 2324 & $\times$ & $\times$ \\ [0.3ex]
$D_{30}(\Delta\Delta)$ & 13150 & 13110 & 3704 & 3526 & 2479 & 2675 \\ [0.3ex]
$D_{12}(N\Delta)$ & 4102 & 4098 & 2453 & 2435 & 2159 & 2232 \\ [0.3ex]
$D_{21}(N\Delta)$ & 8172 & 8155 & 2923 & 2835 & 2219 & 2335 \\ [0.3ex]
\hline\hline
\end{tabular}
\label{tab:dibaryonres}
\end{table}

\subsection{Discussion on results}

We start by examining Tab.~\ref{tab:fitparam}, where the values of the minimized $\chi^2$-function and the values of the phenomenological model parameters for the six different fits are presented. One striking feature is the difference between the minimized values for the primed and the unprimed fits. The unprimed fits ${\bf I}$, ${\bf II}$, and ${\bf III}$ have $\chi^2 \sim 10^3 - 10^5$, which is very large and indicative of a poor fit, while the primed fits ${\bf I}'$, ${\bf II}'$, and ${\bf III}'$ have $\chi^2 \sim 10^0 - 10^1$, suggesting that much better parameter values have been found. This is thought to arise mainly due to $N$ and $NN$ having significantly smaller experimental uncertainties ($\sim 10^{-6}$~MeV) than all other included baryons, making their contributions dominate the $\chi^2$-function and hence giving a larger discrepancy for the other data points. Behavior like this is precisely the motivation behind also performing a 1~\% error fit in the first place. Concerning the parameter values, we note that results for the fits ${\bf I}$ and ${\bf I}'$ deviate significantly from the rest of the fits. For instance, the value of the parameter $M_0$ is much smaller, about 280~MeV, compared to all other fits having $M_0 > 950$~MeV. This is unexpected, since $M_0$ in a sense should represent the total mass of the constituents. Moreover, the parameters $A$ and $D$ -- coefficients for the spin and Casimir eigenvalue terms, respectively -- are radically different for the first two fits. For fits ${\bf I}$ and ${\bf I}'$, $A$ is negative, about $-280$~MeV, while for the rest of the fits, $A$ is positive and of smaller magnitude, about $20$~MeV for fits ${\bf II}$ and ${\bf II}'$ and $(40-50)$~MeV for fits ${\bf III}$ and ${\bf III}'$. The parameter $D$ also differs widely, about 350~MeV for fits ${\bf I}$ and ${\bf I}'$, compared to $(40-50)$~MeV for fits~${\bf II}$ and ${\bf II}'$ and $(10-20)$~MeV for fits~${\bf III}$ and ${\bf III}'$. Finally, the parameters $B$ and $C$ -- coefficients for the hypercharge and isospin terms -- obtain similar values for all six fits, i.e.~$B \in (-200,-190)$~MeV and $C \in (33,41)$~MeV. For reference, we note that the four fits that yield more comparable results (i.e.~fits~${\bf II}$, ${\bf II}'$, ${\bf III}$, and ${\bf III}'$) agree better with the results in Ref.~\cite{Huang:2018izl} than the other two fits (i.e.~fits~${\bf I}$ and ${\bf I}'$), although their method for obtaining the parameter values is not mentioned.

We observe that the results of fits ${\bf I}$ and ${\bf I}'$, using only non-charmed baryons, deviate significantly from the remaining four fits. Since the parameter values obtained for these fits were found to be far off, this is not entirely unexpected. Although most fits follow a general trend for the masses, with $D_{30}$ and $D_{21}$ being the two heaviest, both fits ${\bf I}$ and ${\bf I}'$ show a much more extreme spread with practically all predicted values being rendered unrealistic. However, the remaining fits show a more reasonable behavior, regarding both the primed and the unprimed ones. One explanation for this could be that the heavier charmed baryons, contributing with larger weight to the $\chi^2$-function, result in more consistent parameter values. This might indicate that including charmed baryons is beneficial, provided that the result from data set ${\bf II}$ reproduces the known dibaryon candidates well. In particular, looking at fit ${\bf II}'$, we note that the predicted value for the mass of the $D_{03}$ is about 2324~MeV, which is, in fact, rather close to the experimental value of the mass for the $d^*(2380)$, and basically within our 1~\% error margin for the primed fits. However, the prediction for the $D_{01}$, around 2100~MeV, is too large compared to the experimental value of $NN$ at 1876~MeV. Note that none of the fits were able to describe the deuteron to within even 200~MeV precision.

\begin{figure}
\includegraphics[scale=1]{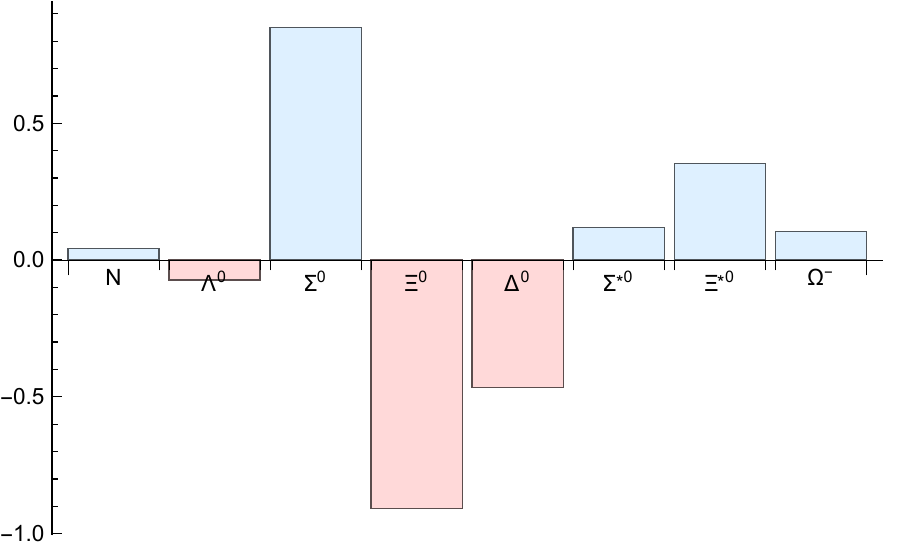}
\caption{Pulls for the eight fitted baryon masses using the values of the free parameters from fit~${\bf I}'$.}
\label{fig:pullsIp}
\end{figure}

\begin{figure}
\includegraphics[scale=1]{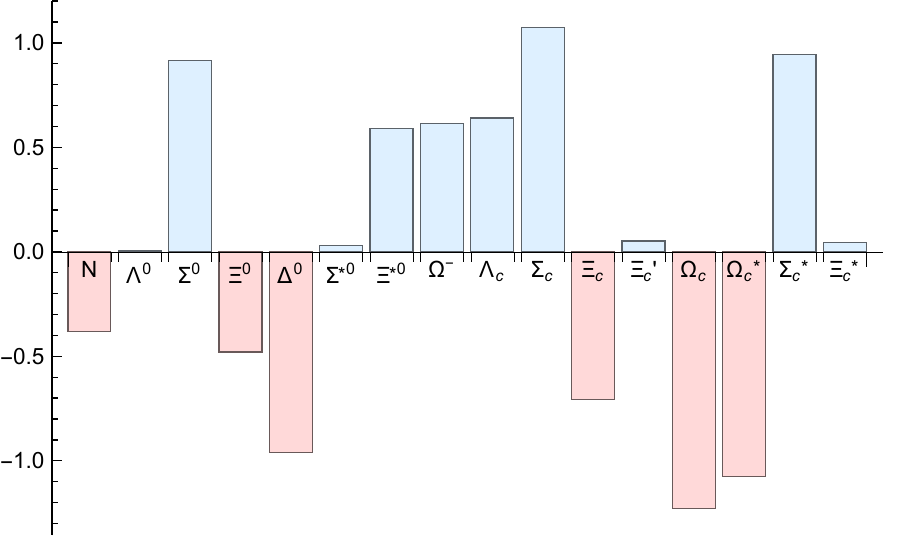}
\caption{Pulls for the 16 fitted baryon masses using the values of the free parameters from fit~${\bf II}'$.}
\label{fig:pullsIIp}
\end{figure}

\begin{figure}
\includegraphics[scale=1]{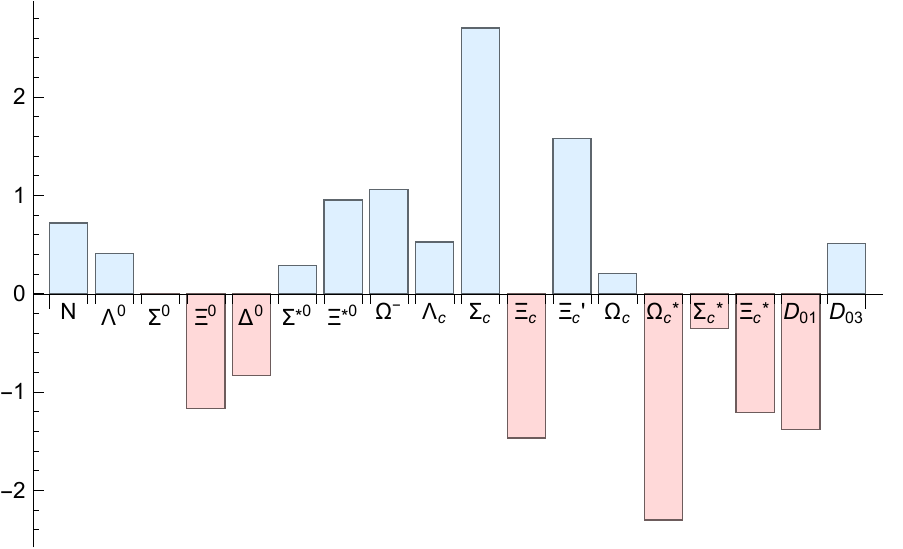}
\caption{Pulls for the 18 fitted masses (16 baryon masses and two dibaryon masses) using the values of the free parameters from fit~${\bf III}'$.}
\label{fig:pullsIIIp}
\end{figure}

One way to investigate the validity of the model is to look at how well it reproduces already known results and data. Since there are only two experimentally determined masses for dibaryon candidates to compare with, we instead look at the well-known baryon spectrum. Figures~\ref{fig:pullsIp}--\ref{fig:pullsIIIp} show the pulls of the calculated baryon masses for the primed fits of the data sets ${\bf I}$, ${\bf II}$, and ${\bf III}$, i.e.~the deviation from the experimental values divided by the uncertainty.
On a general level, we make one immediate observation. The pulls for all of the primed fits are of order unity for data sets ${\bf I}$, ${\bf II}$, and ${\bf III}$, while the individual pulls of the unprimed fits can be as large as $10^3$ (but not presented in Figs.~\ref{fig:pullsIp}--\ref{fig:pullsIIIp}). This agrees with the previously mentioned discrepancies for $\chi^2$, giving an overall better fit at the expense of losing precision for the baryons that have the smallest uncertainties. Another noticeable feature, perhaps not unexpected, is that the data sets including charmed baryons and dibaryon candidates, in general, yield larger pulls for the non-charmed baryons. One explanation for this is that the charmed baryons are heavier, due to the charm quark being more massive than the up, down, and strange quarks, and hence shift the spectra towards larger mass. The same logic applies to the dibaryon candidates, naturally being more massive than conventional baryons. One could therefore argue that the inclusion of charmed baryons contributes to making the parameter values more universal. Although it should be emphasized that the approach of using the GR mass formula with 1~\% error is far from the most precise method conceivable, it is still evident that the basic features of the baryon spectrum are well described by this model. The primed fits give pulls in the range $\pm 3$ for all baryons as well as $D_{01}$ and $D_{03}$, which is of the order one per mille.

\subsection{Comparison with other works}

Since the approach of this study is inspired by the work of Huang {\it et al.}~in Ref.~\cite{Huang:2018izl}, it is reasonable to compare our results to theirs. When fixing the parameters of the GR mass formula, they use experimental masses of the same non-charmed baryons that we present in the upper half of Tab.~\ref{tab:baryon} as well as the dibaryons listed in Tab.~\ref{tab:dibaryon}. In addition to $D_{01}$ and $D_{03}$ with masses assumed to be 1876~MeV and 2380~MeV, respectively, they have also included masses of $D_{10}$, $D_{12}$, and $D_{21}$, albeit followed by a question mark (presumably to signal that these masses are not experimentally verified). Since there is no clear reference to where the values originate, we have chosen to leave them out of our analysis. In any case, the parameter values obtained in their work are (in terms of our notation): $M_0 \simeq 1026.2$~MeV, $A \simeq 56.883$~MeV, $B \simeq -194.70$~MeV, $C \simeq 33.218$~MeV, and $D \simeq 9.4085$~MeV \cite{Huang:2018izl}. We shall also note that they obtain a separate $M_0 \simeq 2091.9$~MeV for dibaryons, whereas we have simply used $2 M_0$ instead. The values of these parameters are in decent agreement with our obtained values for fits ${\bf II}$--${\bf III}'$, i.e.~in terms of sign and order of magnitude.

Regarding the mass spectra, our results are generally larger than those obtained in Ref.~\cite{Huang:2018izl}. For the masses, they obtain 1883~MeV for $D_{10}$, 2394~MeV for $D_{30}$, 2168~MeV for $D_{12}$, and 2182~MeV for $D_{21}$. This is not too surprising, since, as we have mentioned, inclusion of heavier charmed baryons in the data sets naturally should shift the mass spectrum somewhat. However, our results do agree to the extent that $D_{IJ}$ is heavier than $D_{JI}$ for $I > J$ in all cases, meaning that forming bound states of $D_{01}$, $D_{03}$, and $D_{12}$ would be preferred. They present a threshold for the $D_{21}$ being a bound state at about 2171~MeV, which is lower than the value they obtain using the GR approach and also lower than what we obtain. This would indicate that no such state is possible. Further, their more extensive dynamical calculations yield concordant results, after which it is concluded that a bound state $D_{21}$ cannot be obtained within these models, while at the same time accounting for the experimentally observed $d^*(2380)$.

\section{Summary and conclusions}
\label{sec:summary}

Inspired by other works, we have used an extension of the G{\"u}rsey--Radicati mass formula, presented in Ref.~\cite{Santopinto:2016pkp}, to predict the mass spectrum for some non-strange dibaryon candidates. Numerical fits of parameters have been performed based on three different data sets, using both experimental and 1~\% uncertainties for each set, resulting in a total of six fits.

Using only non-charmed baryon masses as input, very large fluctuations in the dibaryon mass spectra have been observed. The other fits, including charmed baryon and non-strange dibaryon masses, have given a more coherent prediction with some results very close to the experimentally observed mass of the hexaquark candidate $d^*(2380)$. The best prediction is 2324~MeV for the mass of $D_{03}(\Delta\Delta)$, whereas 2100~MeV for the mass of $D_{01}(NN)$. Hence, it is not possible to obtain the mass of the deuteron to a better precision than 200~MeV. In general, the fits that use experimental uncertainties have resulted in very large residual values, i.e.~$\chi^2 \sim 10^3 - 10^5$, while the 1~\% error fits have given $\chi^2 \sim 10^0 - 10^1$. For all fits, including charmed baryon and non-strange dibaryon masses as input, all predicted masses of dibaryon states are in the region $(1900,3700)$~MeV.

\vfill

\begin{acknowledgments}
T.O.~acknowledges support by the Swedish Research Council (Vetenskapsr{\aa}det) through Contract No.~2017-03934.
\end{acknowledgments}

\end{document}